\documentstyle[12pt]{article}

\def\be{\begin{equation}}
\def\ee{\end{equation}}

\def\ba{\begin{array}{c}}
\def\ea{\end{array}}

\begin{document}

\titlepage
\vspace*{1cm}

\begin{center}{\Large \bf
 NUMERICAL CONSTRUCTIONS OF PERTURBATION SERIES IN
QUANTUM MECHANICS
 }\end{center}

\vspace{10mm}

\begin{center}
Miloslav Znojil \vspace{3mm}

\'{U}stav jadern\'e fyziky AV \v{C}R, 250 68 \v{R}e\v{z}, Czech
Republic\\

e-mail: znojil@ujf.cas.cz, \today

\end{center}

\vspace{5mm}

\section*{Abstract}

Our several recent improvements of the current textbook (so called
Rayleigh-Schr\"{o}dinger) perturbative representation of bound
states are reviewed. They were all inspired by an adaptive
re-split $H(\lambda)=H^{(0)}+\lambda \,H^{(1)}$ of the
Hamiltonian. Their feasibility is facilitated by the use of
nonstandard bases and by the flexibility of normalization of the
wave functions $\psi(\lambda)$.

\newpage

\section{Introduction}

Anharmonic oscillators often serve as a methodical laboratory in
quantum mechanics \cite{Kunihiro}. Schr\"{o}dinger equation which
determines their bound states is, in the time-independent case, an
elliptic partial differential boundary-value problem, currently
analysed by the variational or perturbation techniques. In the
latter context, one of the most transparent illustrations of the
implementation and of the efficiency of the method is provided by
the elementary one-dimensional quartic example
  \be
 - {d^2 \over d x^2}\, \psi^{}(x) + V^{}(x)\,
\psi^{}(x) =E^{}\, \psi^{}(x)\  \label{SEo}
 \ee
with boundary conditions $\psi(\pm \infty) = 0$, one-parametric
family of potentials $V(x)=x^2+\lambda\,x^4$ and with the so
called Rayleigh-Schr\"{o}dinger formal-series solution
 \be
\ba \psi (x) = \psi^{(0)}(x) + \lambda \ \psi^{(1)}(x) + \lambda^2
\psi^{(2)}(x) + \ldots,
\\
 E = E^{(0)} + \lambda E^{(1)} + \lambda^2 E^{(2)} + \ldots\ .
\ea \label{7}
 \ee
The example nicely illustrates the main merits of the
Rayleigh-Schr\"{o}dinger approach (feasibility, numerical
efficiency near $\lambda \approx 0$) as well as several of its
most characteristic difficulties.

In particular, the mere asymptotic and, hence, divergent-series
character of the anharmonic-oscillator ``solutions" (\ref{7})
divided the physics community into supporters and opponents of the
perturbative considerations. During the last thirty years of
development, the former group definitely prevailed. In what
follows, an illustration of the latter tendency will be presented
via a review of some of a few of my own recent innovations of the
general perturbation formalism.

\section{Formalisms which do not need the unperturbed basis}

One of the most important merits of perturbative considerations
lies in their analytic treatment of certain parameters and in
their enormous (though not always fully appreciated) formal
flexibility. In this sense, there exists an ample variety of
alternatives to the standard textbook representation of wave
functions (\ref{7}) in the (orthonormalized) basis $\{ \,
|n\rangle\,\}$ of the $\lambda=0$ eigenstates of $H(0)$,
\be
 \psi^{(k)}(x)
= \sum_{n=0}^\infty\, \langle x|n\rangle\, p_n^{(k)}\ .
  \label{7bas}
 \ee
In this review of my own contributions to this effort I would like
to mention the representations of wave functions

\begin{itemize}

\item
employing the Runge-Kutta-like discretizations of the domain of
coordinates, both in one \cite{lattice} and more \cite{bilattice}
spatial dimensions;

\item
using the bases with nontrivial orthogonality: Sturmianic
\cite{Sturmians}, complexified \cite{ix3} etc;

\item
switching to techniques which do not need any scalar product at
all and comprise the Taylor-like expansions in the origin
\cite{Taylor} as well as near infinity \cite{Jost}.

\end{itemize}

\noindent All these alterations of eq. (\ref{7bas}) are able to
simplify, first of all, the evaluation of the matrix elements of
the perturbation Hamiltonians $H^{(k)}$.

In this context one should also mention a closely related trick of
working with the so called ``whacko", order-dependent auxiliary
Hamiltonians and bases as employed, e.g., in the non-linear
context of the so called delta-expansions \cite{delta}.

\section{Formalisms with non-diagonal propagators}

In the purely technical sense, the appeal of the textbook
perturbative considerations may be traced back to their frequent
use of the fully diagonalized zero-order models $H^{(0)}$. In
practice, an availability of such a type of an approximant is in
fact very rare. At the same time, people are often seduced by the
extremely transparent formulae obtainable due to the diagonality
of the necessary (so called unperturbed) auxiliary propagators $R$
(schematically: $R \sim (H^{(0)} + const)^{-1}$).

Here, we would like to emphasize, as one of the basic ideas of the
present review, that the construction of the perturbation series
(\ref{7}) does not necessarily require a {\em complete}
solvability/solution of the Schr\"{o}dinger equation
\be
H(\lambda)\,|\psi(\lambda)\rangle
=E(\lambda)\,|\psi(\lambda)\rangle \label{SE}
 \ee
at {\em any particular} (usually, zero) value of the auxiliary
parameter $\lambda=\lambda^{(0)}$.

There exists, in fact, a balance between the strongly redundant
\cite{Skala} requirements of a {\em complete} solution of the
zero-order problem
\be
(H(\lambda^{(0)})-E(\lambda^{(0)}))|\psi(\lambda^{(0)})\rangle=0
\label{zeroorder}
 \ee
and the related enhancement of economy of the solution of the
first-order problem
\be
(H(\lambda^{(0)})-E(\lambda^{(0)}))|\psi^{(1)}\rangle+
(H^{(1)}-E^{(1)})|\psi(\lambda^{(0)})\rangle=0
 \ee
and of its further higher-order descendants
\be
(H(\lambda^{(0)})-E(\lambda^{(0)}))|\psi^{(k)}\rangle=
E^{(k)}|\psi(\lambda^{(0)})\rangle + |\tau\rangle .
 \label{higherorder}
 \ee
One has to defend a compromise and insist on the obvious
observation that the inter-related pair (or rather multiplet of
the $K+1$ linear operator equations) (\ref{zeroorder}) and
(\ref{higherorder}), i.e., the set of the linear operator
equations
\be
A(E^{(0)}) \vec{x} = 0, \ \ \ \ \ \ \ \ A(E^{(0)}) \vec{y}^{(k)} =
\vec{b}^{(k-1)}(E^{(k)}), \ \ \ \ \ \  k = 1, 2, \ldots, K
\label{set}
 \ee
should be considered as a single, unseparated (and, in general,
purely numerical) problem of determination of the unknown
quantities $\vec{x}$, $E^{(k-1)}$ and  $ \vec{y}^{(k)}$.

In this sense, one can accept various computational strategies. In
a way depending on our deeper insight in the structure of the
operator $A$ and in its explicit computational representation, one
could propose the use of some ``surviving" simplifications
offered, e.g.,

\begin{itemize}

\item
by any form of an {\em incomplete} analytic solvability of the
zero order equation (\ref{zeroorder}); whenever available, such a
solvability is still able to simplify the computations
significantly \cite{QES};

\item
by any algebraization of the Schr\"{o}dinger eq. (\ref{SE}); it
provides a new origin of perturbative constructions as exemplified
by the non-variational recipe of ref. \cite{Hill} or, in the much
more numerical context, by the algorithms of ref. \cite{Lanczos}.

\end{itemize}

\noindent In all the above-mentioned examples a key role is played
by the band-matrix structure of the unperturbed Hamiltonian. Also
the sparse-matrix structure of its perturbative ``user-friendly"
components can make the resulting solution of the whole set of
equations (\ref{set}) very straightforward and comfortably
feasible. Such a construction proceeds in fact along the same
lines (and with almost the same ease) as in our introductory
anharmonic example, {\em in spite of the fact} that the
unperturbed propagators themselves remain significantly
off-diagonal.

\section{Working without the model space projectors}

The main price we have to pay for the enhanced flexibility of the
above innovated algorithms is that we lose an immediate insight in
the structure of the unperturbed problem itself.

Several tricks may help to elucidate this structure, at least,
indirectly:

\begin{itemize}

\item
New attention can be paid to the large but still rigorously
terminating cases where, i.a., a new, triple-series formalism has
been shown manageable \cite{triple};

\item
More attention can be paid to the optimization of our choice of
the zero-order Hamiltonian operator itself. For this purpose, the
above-mentioned Runge-Kutta perturbation theory has been combined
with an iterative improvement of $H^{(0)}$, e.g., in ref.
\cite{asy}.

\item In an active manner, we may get even much further. Thus,
controling the mutual methodically relevant relationship between
the feasibility of solution of the homogeneous and non-homogeneous
parts of the Rayleigh-Schr\"{o}dinger hierarchy of equations
(\ref{set}) we can introduce also an {\it ad hoc},
Hartree-Fock-style re-arrangements of many error terms, etc. More
technical details about this efficient strategy can be found,
e.g., in ref. \cite{new} as well as in the further references
listed therein.

\end{itemize}

\noindent One of the main merits of our various modified
perturbation prescriptions may be seen in their shared enhanced
flexibility and in their consequently recurrent character.
Nevertheless, the related algorithms minimizing the distance
between $H(\lambda)$ and $H^{(0)}$ also exhibit a serious weak
point: By their construction, they would all fail in the
(quasi-)degenerate systems. In this context, the most promising
development seems to be represented by the following two different
strategies.

\begin{itemize}

\item
One can get rid of the model space projectors completely. The
first attempt in this direction is provided by an
inverse-iteration-inspired new scheme of ref. \cite{numer}.

\item
One can employ the shooting-like philosophy as implemented in full
detail in ref. \cite{match}.

\end{itemize}

\noindent Technically,  a key problem lies in a preservation of
feasibility of evaluation of the separate Rayleigh-Schr\"{o}dinger
perturbation coefficients.  In this setting the relaxation of the
usual orthogonality of the bases and Hermiticity of the
Hamiltonians seems to open an access to many new unperturbed
$H^{(0)}$. Simultaneously, a weakening and reduction of the
traditional exact solvability requirements opens an access to many
new partially or numerically solvable zero-order approximants with
an immediate approximative role in realistic models.

\section{Outlook}

The main common purpose of the present proposal of transition to
the semi- or numerical evaluation of corrections lies in the
related possibility of a consequent minimization of the absolute
magnitude of the perturbation corrections themselves. In this
sense, the ``size" (whatever it means) of the perturbation
Hamiltonian is efficiently suppressed via a weakening of the
technical constraints imposed upon its eligible zero-order
partners.

We may summarize that the measurable characteristics of various
systems in nuclear and atomic physics or quantum chemistry may be
fairly well approximated using some of the reviewed non-standard
perturbation methods. They reach beyond the vicinity of the usual
exactly solvable zero order models. This significantly extends the
domain of applicability of perturbation expansions to many new
interesting and comparatively complicated systems.

\vskip 0.5cm

\section*{Acknowledgements}

Invited talk presented during the ``Ninth International Colloquium
on
 Numerical Analysis and Computer Science with
Applications" held in Plovdiv, Bulgaria during August 12 - 17,
2000. Partially supported by the organizers and by the grant Nr. A
1048004 of the Grant Agency of the Academy of Sciences of the
Czech Republic.

\vspace{5mm}

\section*{[KEYWORDS]}

\noindent [Ordinary differential equations, partial differential
equations, Schr\"{o}dinger equation, Rayleigh-Schr\"{o}dinger
perturbation theory, bound-state energies $E(\lambda)$,
accelerated convergence, linear algebraic equations for
corrections, applications in physics and quantum chemistry ]

 \vskip 0.5cm

\section*{[AMS 1991 Mathematics Subject Classification]}

81Q05 65F15 47A55 35J10 34A50 15A06


\newpage

\begin{thebibliography}{00}


\bibitem{Kunihiro}
cf. T. Kunihiro, ``Renormalization-group resummation of a
divergent series of the perturbative wave functions of the quantum
anharmonic oscillator,"
 Phys. Rev. D 57 (1998) R2035-9, with further
references.

\bibitem{lattice}
M. Znojil, ``One-dimensional Schr\"{o}dinger equation and its
`exact' representation on a discrete lattice," Phys. Lett. A 223
(1996) 411-6.


\bibitem{bilattice}
M. Znojil, ``A quick perturbative method for Schr\"{o}dinger
equations," J. Phys. A: Math. Gen. 30 (1997) 8771 - 83.




\bibitem{Sturmians}
M. Znojil, ``$r^D$ oscillators with arbitrary $D>0$ and
perturbation expansions with Sturmians," J. Math. Phys. 38 (1997)
5087-97.


\bibitem{ix3}
F. M. Fern\'{a}ndez, R. Guardiola, J. Ros and M. Znojil,
``Strong-coupling expansions for ${\cal PT}$-symmetric oscillators
$V(r) = a\,i\,x+ b\,(ix)^2+ c\,(ix)^3$," J. Phys. A: Math. Gen. 31
(1998) 10105-10112.

\bibitem{Taylor}
M. Znojil, ``Double well model $V(r) = ar^2+br^4+cr^6$ with $a <
0$ and perturbation method with triangular propagators," Phys.
Lett. A 222 (1996) 291-8.


\bibitem{Jost}
M. Znojil, ``Perturbed Poeschl-Teller oscillators," Phys. Lett. A
266 (2000) 254 - 259.

\bibitem{delta}
M. Znojil, ``Three-point Pad\'{e} resummation for anharmonic
oscillators," Phys. Lett. A 177 (1993) 111-20.


\bibitem{Skala}
M. Znojil, ``Comment on the letter ``A new efficient method \ldots
" by L. Sk\'{a}la and J. \v{C}\'{\i}\v{z}ek," J. Phys. A: Math.
Gen. 29 (1996) 5253 - 5256.

\bibitem{QES}
M. Znojil, ``The exact bound-state Ansaetze as zero-order
approximants in
    perturbation theory
 I:   Cz. J. Phys. 41 (1991) 397-408 and
    II:
    Cz. J. Phys. 41 (1991) 497-512.


\bibitem{Hill}
M. Znojil, ``The perturbative method of Hill determinants,"
    Phys. Lett. A 150 (1990) 67-69.



\bibitem{Lanczos}
M. Znojil, ``A perturbative Lanczos method,"
    Phys. Lett. A 155 (1991) 87-93 and

``Perturbation theory for quantum mechanics in its
Hessenberg-matrix representation," Int. J. Mod. Phys. A 12 (1997)
299-304.


\bibitem{triple}
M. Znojil, ``Bound states in the Kratzer plus polynomial
potentials and the new version of perturbation theory," J. Math.
Chem.  26 (1999) 157 - 172.


\bibitem{asy}
M. Znojil, ``Asymmetric bound states via the quadrupled
Schr\"{o}dinger equation," Phys. Lett. A 230 (1997) 283 - 287.


\bibitem{new}
M. Znojil, ``A new form of re-arrangement of the
Rayleigh-Schr\"{o}dinger
 perturbation series,"
Czech. J. Phys. 44 (1994) 545-556.


\bibitem{numer}
M. Znojil. ``Numerically inspired new version of the degenerate
    Rayleigh-Schr\"{o}dinger perturbation theory,"
        Czech. J. Phys. B.40 (1990) 1065-1078.




\bibitem{match}
M. Znojil, ``New perturbation method with the matching of wave
functions," Int. J. Quant. Chem. 79 (2000) 235 - 242.

\end{thebibliography}
\end{document}